# Active thermal metasurfaces for remote heating/cooling by mimicking negative thermal conductivity

Yichao Liu, Kun Chao, Fei Sun*, Shaojie Chen, Hongtao Dai, Hanchuan Chen

*Key Lab of Advanced Transducers and Intelligent Control System, Ministry of Education and Shanxi Province, College of Physics and Optoelectronics, Taiyuan University of Technology, Taiyuan, 030024 China*

\* Corresponding author: sunfei@tyut.edu.cn

**Abstract：**

Remote temperature control can be obtained by a long-focus thermal lens that can focus heat fluxes into a spot far away from the back surface of the lens and create a virtual thermal source/sink in the background material, around which the temperature field distribution can be remotely controlled by changing the parameters of the thermal lens. However, due to the lack of negative thermal conductivity, the existing thermal lenses have extremely short focal lengths and cannot be used to remotely control the temperature field around the virtual thermal source/sink. In this study, we theoretically propose a general approach to equivalently realize negative thermal conductivity by elaborately distributed active thermal metasurface (ATMS), then use the proposed ATMS to implement a novel thermal lens with long focal length designed by transformation thermodynamics, and experimentally verify the performance of the designed long-focus thermal lens with measured focal length $f$=19.8mm for remote heating/cooling. The proposed method expands the scope of the thermal conductivity and open up new ways to realize unprecedented thermal effects with effective negative thermal conductivity, such as "thermal surface plasmon polaritons", thermal superlens, thermal tunneling effect, and thermal invisible gateway.

Natural materials have a limited range of thermal conductivity, making it difficult to achieve some unprecedented thermal effects. To expand the range of thermal conductivities, some novel thermal materials with artificial structures have been proposed, often referred to as thermal metamaterials and metasurfaces[1-3]. As thermal metamaterials and metasurfaces can achieve more diverse and wider range of thermal conductivity, e.g., inhomogeneous anisotropic values, tunable thermal conductivity, and even a thermal conductivity tending to infinity[4], the capability to control temperature fields has been improved drastically, such as achieving thermal cloaking[5-9], thermal camouflage [10-16], thermal concentrator[17], thermal rectification and thermal diodes[18-21], thermal Hall effect[22-24], thermal encoding[25], thermal buffering[26], and thermal lens[27-30]. These findings broaden the scope of thermodynamic research and pave the way for the realization of unusual thermal conductivity values. However, when compared to the capability of controlling electromagnetic waves, the current control capability for temperature fields can be further improved. For example, it is simple to realize a long-focus lens that can focus an electromagnetic wave into a spot that appears to be created by a virtual electromagnetic source in free space. Nevertheless, theoretical design and experimental demonstration of a long-focus thermal lens, which can focus temperature fields into a spot by thermal conduction and make it look like the temperature field generated by a virtual temperature source/sink in the background material, are still challenging. The physical mechanism for this difficulty lies in the fact that the second law of thermodynamics only allows heat transporting from high temperature region to low temperature region spontaneously, implying that long-focus thermal lenses can only be achieved with active materials (e.g., negative thermal conductive materials).

From the perspective of transformation optics[31,32] and transformation thermodynamics[33], there is no intrinsic difference between the focusing of electromagnetic fields and the focusing of temperature fields, which can both be treated as creating a virtual electromagnetic or thermal source/sink by translation or folding coordinate transformation[34,35]. There have been several interesting designs on thermal focusing by transformation thermodynamics, such as producing thermal focal spots with predesigned shape[36] and position[27-30,34], and even creating multiple focal spots from a single thermal source[37,38]. However, the focal length of these thermal lenses is often extremely short, i.e., the distance between the virtual source (focal spot) and the back surface of the lens is close to zero in Fig. 1**a**, which cannot be used for remote control of temperature fields. Although a long-focus thermal lens can be designed by a folding transformation in transformation thermodynamics, where the virtual thermal source/sink is far away from the back surface of the lens in Fig. 1**b**, this will inevitably introduce a negative thermal conductive material that doesn't exist in nature. Although there are some theoretical studies on negative thermal conductive materials, i.e., how theoretically use negative thermal conductivity to achieve various novel thermal phenomena such as thermal cloaking[39], golden touch phenomenon[40], and remote heating/cooling effect[35], however, there is still a lack of research on how to realize negative thermal conductive materials by thermal metamaterials/metasurfaces. Unlike electromagnetic metamaterials, where negative permittivity or permeability can be obtained by resonant artificial units, thermal conduction does not have a corresponding resonance effect, and therefore negative thermal conductive materials cannot be designed by analogy with electromagnetic resonant metamaterials of negative permittivity/permeability.

**Design concept**

To equivalently realize negative thermal conductive materials, which can then be used to construct long-focus thermal lens for remote heating/cooling, we propose a general method on how to place heat sources with pre-designed continuous power distribution along the boundary of the hypothetical negative thermal conductive material, which can achieve the same temperature field modulation effect in positive thermal conductive materials as in negative thermal conductive materials. Then, we study how to realize the boundary heat power distribution by finite active thermal metasurfaces (ATMS), which can create special temperature distributions that previously had to be achieved with negative thermal conductive material, such as long-focus thermal focusing in Fig. 1**b** and 1**c**. As an example, we design a long-focus thermal lens with negative thermal conductivity using transformation thermodynamics to remotely control thermal fields (see Fig. 1**b**), and use the proposed method to equivalently realize the designed long-focus thermal lens by ATMS without any negative thermal conductive material (see Fig. 1**c**). The long-focus thermal focusing capability for remote heating\cooling is verified by both numerical simulation and experimental measurement (a fabricated thermal lens with focal length $f = 19.8$mm). The proposed method provides how to achieve the same temperature field control effect in negative thermal conductive materials by placing ATMS with pre-designed heat power distribution in positive thermal conductive materials, thus extending the range of thermal conductivity that can be achieved by current thermal metamaterials/metasurfaces. This work makes up for the inability to mimic some interesting optical phenomenon in thermotic due to the lack of negative thermal conductivity, such as superlens[41], invisible gateway[42], tunneling effect[43], and surface plasmon polaritons[44].

**Equivalence between negative thermal conductive material and ATMS**
For the two-dimensional (2D) case, assuming a piece of material with negative thermal conductivity $-\kappa_a$ ($\kappa_a > 0$, the blue region 1 in Fig. 2**a**) is embedded in the background material with positive thermal conductivity $\kappa_b$ ($\kappa_b > 0$, the white region 0 in Fig. 2**a**). The heat flux in normal direction should be continuous at each point of the boundary without considering the interface thermal resistance[45,46], i.e.,

$$q_\mathbf{u} = -\kappa_b \mathbf{u} \cdot (\nabla T)_{region0} = \kappa_a \mathbf{u} \cdot (\nabla T)_{region1} , \qquad (1)$$

where **u** is the unit vector normal to the interface (indicated by the red arrow in Figs. 2**a**-2**c**). Due to the continuity of the heat flux as in Eq. (1), the temperature gradients aside the boundary between positive and negative thermal conductive materials have opposite directions in Fig. 2**a**. If the negative thermal conductive material in Fig. 2**a** is replaced by the positive thermal conductivity in Fig. 2**b**, to keep the temperature field distribution in each region (including the boundary) still the same as in Fig. 2**a**, the heat fluxes around the boundary close to the region 1 become $-\kappa_a \mathbf{u} \cdot (\nabla T)_{region1}$ in Fig. 2**b**, which have opposite directions of the heat fluxes at the same position as in Fig. 2**a**. In this case, it is necessary to introduce boundary heat source (i.e., described by the boundary heat power density $q_s$) to keep the continuity of the heat flux at each point on the boundary in normal direction while not having an additional effect on the temperature field in the two regions (i.e., to keep the temperature field inside the region (0) and the region (1) exactly the same in Fig. 2**a** and Fig. 2**b**), which can be given by:

$$q_s = -2q_\mathbf{u} = -2\kappa_a \mathbf{u} \cdot (\nabla T)_{region1} . \qquad (2)$$

The boundary heat power generated per unit area (or per unit length for 2D case) has the same dimension as the heat flux and is related to the normal component of the heat flux at the interface. The minus sign in Eq. (2) indicates that the boundary heat source is used to eliminate the effect of the heat flux from both sides at the boundary, while the factor 2 in Eq. (2) lies in the fact that heat fluxes from two normal directions aside the boundary all contribute to the boundary heat sources. Note the heat flux in tangential direction is continuous and will not contribute to the boundary heat sources.

In order to discretize the boundary heat source $q_s$ in Eq. (2) into a finite number of ATMSs with designed heat power distribution, the boundary is discretized into $M$ curve segments with equal interval $\Delta$. $M$ ATMSs are placed at the center of each curve segment shown in Fig. 2**c**. The heat power of the $m$-th ATMS ($m=1,2,3... M$, and $M$ is the total number of the ATMSs) can be designed as the line integration of the boundary heat source $q_s$ over the $m$-th curve segment:

$$Q_m = \int_{\theta_m - \frac{\Delta\theta_m}{2}}^{\theta_m + \frac{\Delta\theta_m}{2}} q_s r d\theta \ . \tag{3}$$

To initially verify the validity of the discretized ATMSs with designed heat power distribution in Eq. (3) that can equivalently perform as negative thermal conductive material, numerical simulations of a three-layer structure are shown in Figs. 2**d** and 2**e**. Numerical simulation results show that the temperature distributions in all regions are exactly the same when the negative thermal conductive material in the middle layer in Fig. 2**d** is replaced by the positive thermal conductive background material and the designed ATMSs on both sides of the middle layer in Fig. 2**e**, which verifies the equivalence between negative thermal conductivity and the designed ATMSs.

To verify the generality of the proposed method for equivalently achieving negative thermal conductive materials by ATMS, additional simulations are given in Supplementary Note 1 when the boundary between positive and negative thermal conductive materials have arbitrary irregular shapes, which verify the negative thermal conductive material with arbitrary shapes can be equivalently realized by the designed ATMSs in Eq. (3).

**Long-focus thermal lens for remote heating/cooling by transformation thermodynamics**

First, we design a long-focus thermal lens with negative thermal conductivity by transformation thermodynamics to achieve remote heating\cooling effect as shown in Fig. 1**b**. The reference space and the physical space are shown in Figs. 3**a** and 3**b**, respectively, in which the quantities are distinguished by with and without primes. In the reference space, a line heat source with constant power $Q_0'$ is set at $\rho' = h_i$ ($\theta' = 0$) between the circle $C_2$ with radius $R_2$ and the circle $C_3$ with radius $R_3$ ($R_3 > R_2$) in Fig. 3**a**. Next, we will show how to transform the location of the heat source ($\rho' = h_i$, $\theta' = 0$) in the reference space to another location ($\rho = h_o, \theta = 0$) in the physical space by transformation thermodynamics. We fix the boundary of the circle $C_2$ and fold the circle $C_3$ inward to another circle $C_1$ with radius $R_1$ ($R_1 = R_2^2/R_3$), i.e., using the transformation

$\rho = R_2^2/\rho'$ and $\theta = \theta'$ to fold the region $R_2 < \rho' < R_3$ in Fig. 3a to the region $R_1 < \rho < R_2$ in Fig. 3b. To keep the continuity of the space, the region inside the circle $C_3$ should be squeezed to the interior of the circle $C_1$ by the transformation $\rho = \rho' R_1/R_3$ and $\theta = \theta'$. Consequently, the source is squeezed to the new position at $\rho = h_o = h_i R_1/R_3$ in the physical space (the variations of the coordinate grids and the heat source during this process can be found in Supplementary Movie 1). Assuming that a uniform background medium with positive thermal conductivity $\kappa_b$ fills the whole reference space, the 2D in-plane thermal conductivity in each region of the physical space can be calculated by transformation thermodynamic[33] as (see Supplementary Note 2 for detail):

$$\begin{cases} \kappa_1 = \kappa_b & \text{for region } \rho < R_1 \\ \kappa_2 = -\kappa_b & \text{for region } R_1 \leq \rho < R_2 \\ \kappa_3 = \kappa_b & \text{for region } \rho \geq R_2 \end{cases} \quad (4)$$

In the physical space, the thermal material in region $R_1 < \rho < R_2$ has negative thermal conductivity $-\kappa_b$, which corresponds to the designed long-focus thermal lens, while the thermal conductivity of the remaining parts is kept as the uniform background medium with positive thermal conductivity $\kappa_b$. Moreover, the power of the transformed heat source located at ($\rho = h_o = h_i R_1/R_3$, $\theta = 0$) in the physical space can also be calculated by the transformation thermodynamics (see Supplementary Note 2), which is the same as the power of the heat source located at ($\rho' = h_i$, $\theta' = 0$) in the reference space, i.e., $Q_0 = Q_0'$.

From the perspective of transformation thermodynamics, the function of the materials with the thermal conductivities in Eq. (4) is to 'fold' the heat source of power $Q_0$ at ($\rho = h_o$, $\theta = 0$) in the physical space to create a virtual heat source (i.e., a thermal image) at ($\rho = h_i$, $\theta = 0$), which performs as a long-focus thermal lens with the focal length $f = h_i - R_2$ (see Fig. 3d). The temperature field distribution created by this virtual heat source is almost the same as that produced by a heat source of power $Q_0$ placed at ($\rho' = h_i$, $\theta' = 0$) in a uniform positive thermal conductive medium $\kappa_b$ in the reference space (see Fig. 3c) for the region $\rho > R_2$, except the regions near the boundaries between positive and negative thermal conductive materials, where the temperature is extremely high. This interesting phenomenon is similar to the surface plasmon polaritons generated on the boundary between dielectric and metal (negative permittivity) in optics, which can be treated as "thermal surface plasmon polaritons".

Considering the geometric and coordinate transformation relations, the focal length of the designed long-focus thermal lens with negative thermal conductivity in Fig. 3b can be further expressed as

$$f = h_i - R_2 = R_2 \left( \frac{h_o R_2}{R_1^2} - 1 \right). \quad (5)$$

The focal length is independent of the thermal conductivities of the background material and the negative thermal conductive material, and is entirely determined by the geometric parameters of the lens ($R_1$ and $R_2$) and the location of the actual heat source $h_o$. Eq. (5) shows that the longer focal length can be obtained by increasing $R_2$ or reducing $R_1$. If the source position $h_o$ and inner boundary of the lens $R_1$ is fixed, longer focal length can be achieved by a larger outer boundary of the lens $R_2$, (see additional calculated results in Supplementary Note 3). By adjusting the power of the heat source $Q_0$ in the physical space, the temperature field distribution around the virtual source (at $f=h_i-R_2$ from the back surface of the thermal lens) can be changed correspondingly, e.g., the remote heating and cooling can be achieved by setting the heat source to release and absorb heat power, respectively. This remote temperature control capacity will be further studied after introducing how to implement the designed long-focus thermal lens with ATMSs later.

**Long-focus thermal lens by ATMS**

Second, we show how to realize the designed long-focus thermal lens with negative thermal conductivity in Eq. (4) by the proposed ATMS in Eq. (3) to achieve the same remote thermal focus effect in Fig. 3**d**. As the thermal conductivity in Eq. (4) is uniform for each region in the physical space, we can calculate the temperature distribution in each region directly by solving the two-dimensional thermal conduction equation with a line heat source $\nabla(\kappa\nabla T) = A_0\delta(\rho-h_o)\delta(\theta)/\rho$. The temperature inside the designed long-focus thermal lens with negative thermal conductivity can be written as (detail calculations are given in Supplementary Note 3):

$$T = \frac{A_0}{2\pi\kappa_b}\ln\rho + \frac{A_0}{2\pi\kappa_b}\sum_{n=1}^{N}\frac{1}{n}\left(\frac{\rho h_o}{R_1^2}\right)^n \cos(n\theta) + T_b + \frac{A_0}{2\pi\kappa_b}\ln R_b - \frac{A_0}{\pi\kappa_b}\ln R_2, \quad R_1 \leq \rho \leq R_2, \quad (6)$$

where the constant $T_b$ is a constant temperature of the far-field background (i.e., the room/environment temperature), and the positive integer $N$ is the highest order in the Fourier series. Eq. (6) gives the temperature field distribution in the region with negative thermal conductive material, thus the heat flux distribution on the boundaries $C_1$ and $C_2$ can be calculated from the temperature field gradient by Eq. (1), and then the equivalent two-boundary heat sources with continuous thermal flux distribution along circles $C_1$ and $C_2$ can be calculated by Eq. (2), which are given as,

$$\begin{cases} q_{s,R_1} = -\dfrac{A_0}{\pi R_1} - \dfrac{A_0}{\pi R_1}\sum_{n=1}^{N}\left(\dfrac{h_o}{R_1}\right)^n \cos(n\theta) \\ q_{s,R_2} = \dfrac{A_0}{\pi R_2} + \dfrac{A_0}{\pi R_2}\sum_{n=1}^{N}\left(\dfrac{h_o R_2}{R_1^2}\right)^n \cos(n\theta) \end{cases}. \quad (7)$$

The analytical calculations show that the temperature created by two-boundary heat sources along circles $C_1$ and $C_2$ in Eq. (7) together with a line heat source $A_0\delta(\rho-h_o)\delta(\theta)/\rho$ is the same as the temperature created by one-boundary heat source along circle $C_2$ without any line heat source for region $\rho>R_2$, which is given as (see Supplementary Note 4 for detail),

$$\tilde{q}_{s,R_2} = \frac{A_0}{2\pi R_2} + \frac{A_0}{\pi R_2} \sum_{n=1}^{N} \left( \frac{h_o R_2}{R_1^2} \right)^n \cos(n\theta) . \tag{8}$$

Simulated results in Figs. **4a** and **4b** verify that two-boundary heat sources with a line heat source (i.e., $q_{s,R_1}$, $q_{s,R_2}$ in Eq.(7) and $A_0 \delta(\theta)\delta(\rho - h_o)/\rho$) and one-boundary heat source without any line heat source (i.e., $\tilde{q}_{s,R_2}$ in Eq.(8)) can create the same expected remote thermal focus effect, which just like the effect produced by negative thermal conductive materials in Fig. **3d**. Therefore, an expected remote virtual heat source outside the outer boundary $C_2$ can be produced by any set of equivalent boundary heat sources in Eqs. (7) and (8). Similar "thermal surface plasmon polaritons", i.e., oscillatory distribution of high and low temperature fields, occurs near equivalent boundary heat sources, where the magnitude of the temperature field is extremely high around the boundaries.

Next, we discretize the one-boundary heat source with continuous thermal flux distribution $\tilde{q}_{s,R_2}$ in Eq. (8) into $M$ equal spaced ATMSs by Eq. (3), and the heat power of the $m$-th ATMS can be written as,

$$\tilde{Q}_{R_2,m} = \frac{A_0}{2\pi} \Delta\theta + \frac{A_0}{\pi} \sum_{n=1}^{N} \left( \frac{h_o R_2}{R_1^2} \right)^n \frac{2}{n} \sin\left( \frac{n}{2} \Delta\theta \right) \cos(n\theta_m), \tag{9}$$

where $\Delta\theta = 2\pi/M$, $\theta_m$ is the polar angle of the $m$-th ATMS (see Fig. **2c**), and the total number of the ATMS $M$ is designed as 2 times the highest order in the Fourier series $N$, i.e., $M = 2N$. Figures **4c-f** show the simulated temperature distributions, where the numbers of ATMS are chosen as (a) $M = 6$, (b) $M = 12$, (c) $M = 18$, and (d) $M = 24$. The simulated results show that the temperature distribution in the region of concerned tends to be more similar to the temperature distribution created by a virtual source in Fig. **3c** as $M$ increases, while the focal length of the lens remains almost the same as the theoretically expected value in Eq. (5) when $M$ increases to 12. To achieve a satisfied remote heating effect, a larger $M$ is expected. However, as $M$ increases, more ATMS are required to realize the thermal lens and more power will be consumed. Therefore, the thermal lens with $M=12$ ATMS are chosen to be fabricated in the subsequent experiments.

**Experimental measurements**
The schematic diagram of the experimental setup is shown in Fig. **5a**, where a thermal lens with 12 ATMSs is fabricated to experimentally demonstrate the remote heating and cooling effect. The ATMSs with predesigned power can be realized by thermoelectric (TE) components, which is based on the Peltier effect and can release or absorb heat by applying forward or reverse electric currents[47]. The released and absorbed heat power $Q_H$ and $Q_L$ of each ATMS can be calculated as[48,49],

$$\begin{cases} Q_H = \alpha I T_H + \frac{1}{2}I^2 R - \kappa_T \Delta T \\ Q_L = \alpha I T_L - \frac{1}{2}I^2 R - \kappa_T \Delta T \end{cases}, \quad (10)$$

where $I$ is the amplitude of the input current, and $\Delta T = T_H - T_L$ is the temperature difference between the high temperature surface $T_H$ and the low temperature surface $T_L$ of the TE component. $\alpha$, $R$ and $\kappa_T$ are the Seebeck coefficient, electrical resistance, and thermal conductance, respectively, which characterize the TE component.

12 ATMSs (with size of 6.0mm×6.0mm×2.05mm) are arranged as a circle with radius $R_2 = 33$mm, which are sticked to a thick circular copper plate with radius 125mm and thickness 3mm by thermally conductive silicone grease (with thermal conductivity of 6W/m/K). The copper plate serves as a thermal sink to absorb or release heat at the bottom surface of the TE components to best equivalently perform as theoretically required ATMS. Graphite sheets with sizes of 500mm×500mm×0.1mm and lateral thermal conductivity about 1500 W/m/K[50] is covered on the top surface of each ATMS, which serves as the background material. Foam board with thickness of 3mm and thermal conductivity about 0.035W/m/K[51] is placed on the bottom side of the graphite sheets to reduce the thermal convection. More details on sample preparation are given in Supplementary Note 6.

To demonstrate the remote heating effect, the virtual thermal image created by the designed thermal lens should be a high-temperature image (i.e., $A_0 > 0$ in Eq. (9)). In this case, the required thermal powers of the 12 ATMSs can be calculated by Eq. (9) with $A_0 = 0.187$W/m, which are normalized as the power ratio by the power of the first ATMS in the experiment. Once required source power ratio of each ATMS is obtained, the required electric currents applied on each ATMS can be initially estimated by Eq. (10) under the linear approximation, which will be slightly corrected during the experimental testing. Then, a real-time temperature distribution of the detection region with the fabricated thermal lens can be observed by an infrared camera (FOTRIC 288). The general procedure of the experimental measurement can be found in Supplementary Movie 2. More details on measurement process are given in Supplementary Note 7. Figs. 5**b** and 5**c** show the simulated temperature distribution and measured temperature distribution captured by the infrared camera. The measured result consists well with the simulated result, which shows the remote heating effect is realized by a long-focus thermal lens with outer radius $R_2 = 33$mm and focal length $f = 19.8$mm. The real-time measured temperature distribution can be found in Supplementary Movie 3.

For a remote heat cooling effect, the virtual thermal image created by the designed thermal lens will perform as a low-temperature image (i.e., $A_0 < 0$ in Eq. (9)). In this case, the required power ratio of each ATMS and corresponding applied electric currents on each ATMS can be obtained using the same steps (see Supplementary Note 7 for more details). To see the cooling effect clearly, the temperature created by a line high-temperature heat source (with a constant normalized power 0.187W/m and located at $(h_i, 0)$) to be cooled later is simulated in Fig. 5**d** and measured in Fig. 5**e** as a reference.

If the reference high-temperature heat source is set at the focal point of the designed thermal lens, both simulated and measured results show the temperature around the high-temperature heat source is greatly decreased in Figs. 5**f** and 5**g**. The corresponding temperature distributions along the line $x = h_i$ for each case are plotted at the right side for Figs. 5**d-g**. Compared with the reference in Figs. 5**d** and 5**e**, expected heat flux suppressing effect and remote cooling effect are observed with the designed thermal lens in Figs. 5**f** and 5**g**. The temperature drop of the fabricated thermal lens is about 1 degree in the experiment, which can be further improved by increasing the power of each ATMS unit or the number of ATMSs. Note that increasing the power of each ATMS unit would result in a temperature drop in the surrounding area other than the target point, while increasing the number of ATMSs can provide a higher temperature drop at the target point almost without affecting the temperature of the surrounding area.

**Potential applications**
The proposed approach to equivalently realize negative thermal conductive material by ATMS may have many novel applications on temperature control and thermal engineering. In analogy to superlens in optics that can achieve subwavelength resolution by negative refractive index, a thermal superlens, which is a negative thermal conductive slab that can significantly improve the resolution of thermal imaging systems, can be realized by the proposed ATMS. Similar to the surface plasmon polaritons that occur on the interface between positive and negative refractive index media for optical superlens, "thermal surface plasmon polaritons" appear in thermal superlens, which is analyzed in Supplementary Note 8.1. With the help of thermal superlens by the proposed ATMS, the temperature distribution of two heat objects very close to each other can be clearly resolved when the observation plane is far away from the object plane (see more in Supplementary Note 8.2).

Similar to electromagnetic complementary media pair in optics, thermal complementary media pair are two thermal media whose thermal conductivities has the same magnitude but opposite sign. The region occupied by the thermal complementary media pair is equivalent to the void region for heat flow, which can be used to 'eliminate the space' thermally. The proposed ATMS can be utilized to realize thermal complementary medium of the thermal insulation wall, which can be placed aside the thermal insulation wall to enable heat flow to continue propagating through thermal insulation wall and achieve a thermal tunneling effect (more detailed can be found in Supplementary Note 8.3). In this case, the heat flux can pass through the thermal insulation wall, which performs as a thermally conductive but electric non-conductive medium.

In thermal protection applications, it is often desired to prevent external heat flow from entering the region containing some thermally sensitive components, yet allow electric current to enter the same region to power these components, which is difficult to realize by natural materials due to the Wiedemann-Franz Law. A thermal invisible gateway, which can be realized by the designed ATMS, can block the heat flow and allow electric current through the gateway (see more in Supplementary Note 8.4).

From various examples, attributed to the proposed method of implementing negative thermal conductive materials with ATMS in this study, previous novel optical phenomena and electromagnetic devices with various functions based on negative refractive index media, complementary materials and surface plasmon polaritons can be extended to the thermal field correspondingly, thus providing an unprecedented new approach to control temperature field and heat flux.

**Remote temperature control by ATMS**

The proposed long-focus thermal lens by ATMS can provide three ways for remote temperature control, such as switching function between cooling and heating at the thermal focal point, tuning the temperature value at the focal point, and changing the position of the thermal focal spot. These three different functions of remote temperature control can all be achieved real-timely by changing the electric currents loaded on ATMS.

By adjusting the sign and value of the applied current on each ATMS, the amplitude $A_0$ of the virtual heat source in the reference space in Eq. (9) can change its sign or value correspondingly, which in turn can generate a heating/cooling function switch (see Figs. 5**c** and 5**g** for measurement results) or a temperature value adjustment at the thermal focal spot (see more in Supplementary Note 9.1 and Supplementary Movie 4). Note that this real-time remote temperature control effect only changes the temperature in the region near the thermal focal spot (i.e., the blue dots in Fig. 4 and Figs. 5**b-c**), while the temperature in the region outside the thermal focal spot is basically unchanged as the room/environment temperature. This method of remote temperature field control can be used in biochemistry experiments to provide real-time heating/cooling of a specific region (e.g., the region where biochemical reactions occur) while the temperature field in the surrounding area remains the room/environment temperature.

The focal length of the thermal lens $f$ in Eq. (5) can be tuned by changing the parameter of the thermal lens ($R_1$ and $R_2$) or the position of the line thermal source ($h_o$), which can be equivalently achieved by tuning the heat power of each ATMS in Eq. (9) (see Supplementary Note 9.2 and Supplementary Movie 5). The angular position of the focal point can also be tuned by introducing rotation angle parameter in Eq. (9), which can also be realized by tuning the power of each ATMS (see Supplementary Note 9.2 and Supplementary Movies 6-7). Therefore, the position of the thermal focal spot, i.e., both the focal length and angular position, can be changed real-timely by adjusting the applied current on each ATMS. This adjustment of the focal position can be spatial continuous or discrete, which may have applications on spatial alternating heating/cooling, channel assignment of thermal signals, and time division heating.

**Conclusions**

In conclusion, a method to equivalently realize negative thermal conductivity by ATMSs is proposed and used to design a series of novel thermal devices that were previously only available for electromagnetic waves, e.g., thermal 'surface plasmon polaritons', thermal superlens, thermal tunneling effect, and thermal invisible gateway. The proposed method is also utilized to design a long-focus thermal lens, which can focus heat fluxes into a spot far away from the lens for a remote temperature field control. Both simulated and measured results verify the validity of the proposed method and the ability of the long-focus lens for remote temperature control, which includes heating-cooling switching and thermal target shifting (e.g., focal length and orientation). A fabricated thermal lens with focal length $f$=19.8mm for both remote heating and cooling is verified experimentally, where an expected temperature jump/drop at the precise target point is observed. The proposed method extends the scope of thermal conductivity and has potential application in remote temperature control and thermal engineering, e.g., dynamic thermotic control system, thermal imaging, thermal illusions, and thermal coupling/communication.

**Data availability**
The main data and models supporting the findings of this study are available within the paper and Supplementary Information. Further information is available from the corresponding authors upon reasonable request.

**Acknowledgments**
This work is supported by the National Natural Science Foundation of China (Nos. 61905208, 61971300, and 12274317).

**Author contributions**
Yichao Liu: Methodology, Software, Experiment design, Writing original draft.
Kun Chao: Sample preparing, Measurement, Data recording.
Fei Sun: Conceptualization, Supervision, Writing review & editing.
Shaojie Chen: Assist with sample preparation and measurement.
Hongtao Dai: Assist with measurement and video recording.
Hanchuan Chen: Literature investigation, Assist with measurement.
All authors contribute to discussions.

**Competing interests**
The authors declare no competing financial interests.

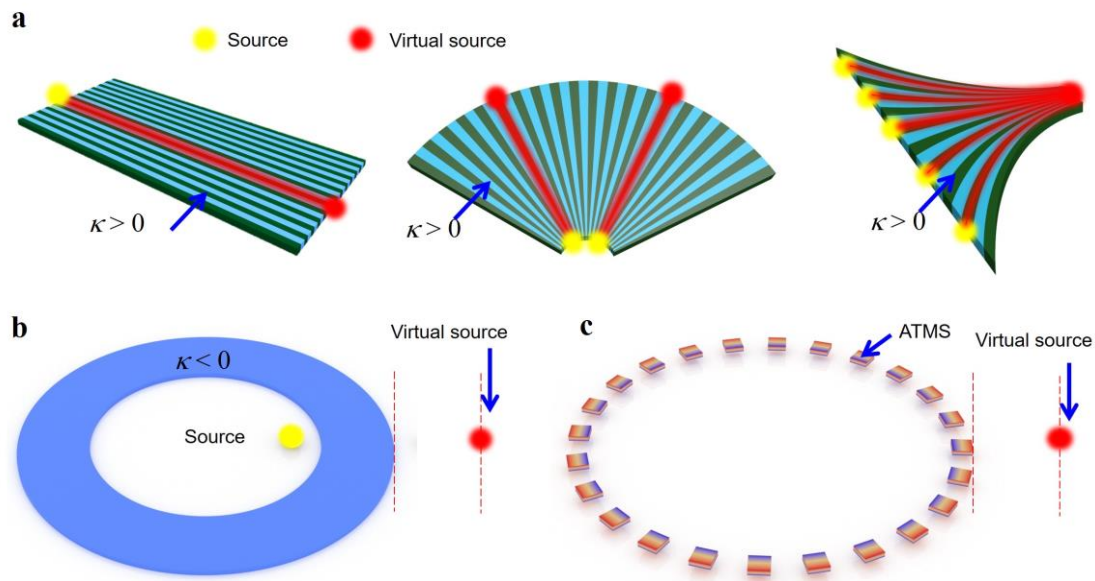

**Fig. 1| Direct control and remote control of temperature fields. a,** Direct control on temperature fields by a short-focus thermal lens with anisotropic thermal conductivity (cyan and green blocks), in which the real thermal sources (yellow dots) and the virtual sources (red dots, i.e., focal spot) are directly linked by the lens. In this case, the distance between the focal spot and the back surface of the lens is zero (i.e., the focal length is zero). Left subplot: a planar plate thermal lens[27]; Middle subplot: a fan-shaped thermal lens[30]; Right subplot: a thermal lens with curved principal axes of the anisotropic thermal conductivity[28]. **b,** Remote control on temperature fields by a long-focus thermal lens with negative thermal conductivity (colored blue): if a real thermal source (yellow dot) is set inside the lens, a thermal focal spot (red dot) that looks like a virtual thermal source will be created outside the lens. **c,** Remote control on temperature fields by a circle enclosed by ATMS (colored rainbow) without any negative thermal conductive material. In this case, a thermal focus spot can be generated without a real thermal source by using well designed boundary heat power distributions of the ATMS. Unlike the short-focus thermal lens in (**a**) where the thermal focus spot occurs at the back surface of the lens, the structures in (**b**) and (**c**) are equivalent to long-focus thermal lenses, which can produce a thermal focus spot (virtual source) located at a certain distance from the back surface of the lens (indicated by the distance between two red dashed lines).

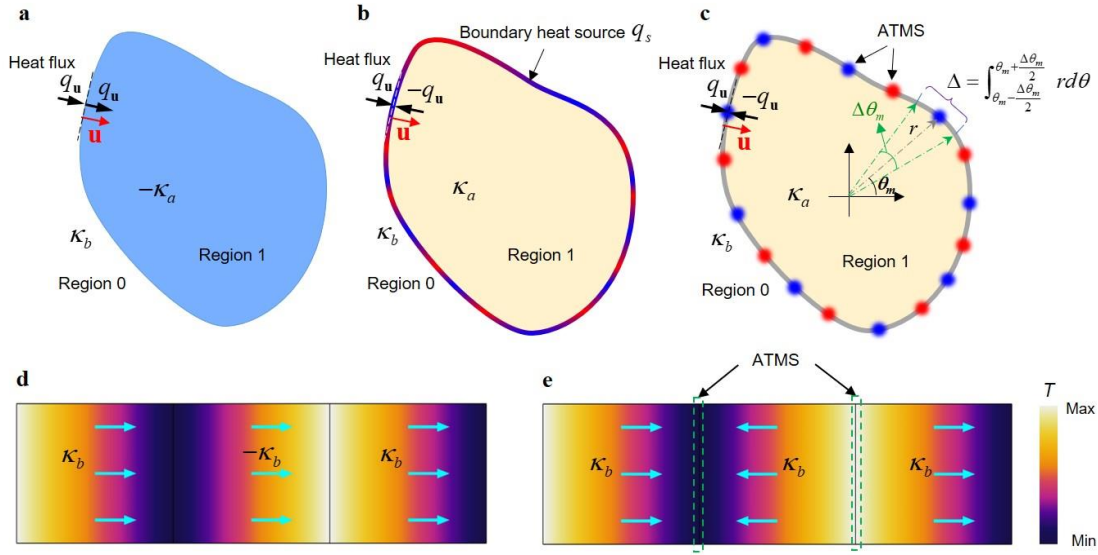

**Fig. 2| Schematic illustration** on how to equivalently realize negative thermal conductive material by the continuous boundary heat sources in Eq. (2) or the discrete ATMSs in Eq. (3). **a**, an arbitrarily shaped negative thermal conductive material embedded in a positive thermal conductive background material. **b,c**, the negative thermal conductive material is replaced by the positive thermal conductive background material surrounded by (**b**) the continuous boundary heat sources and (**c**) the discrete ATMSs. **d,e**, Simulated temperature distributions of a three-layer structure for (**d**) negative thermal conductive material sandwiched between two positive thermal conductive background materials, and (**e**) positive thermal conductive material sandwiched between two positive thermal conductive background materials with two sets of designed ATMSs by Eq. (3) (inside two dashed boxes) on the boundaries when constant high/low temperature sources are applied at the left/right boundaries. Cyan arrows indicate the directions of heat fluxes.

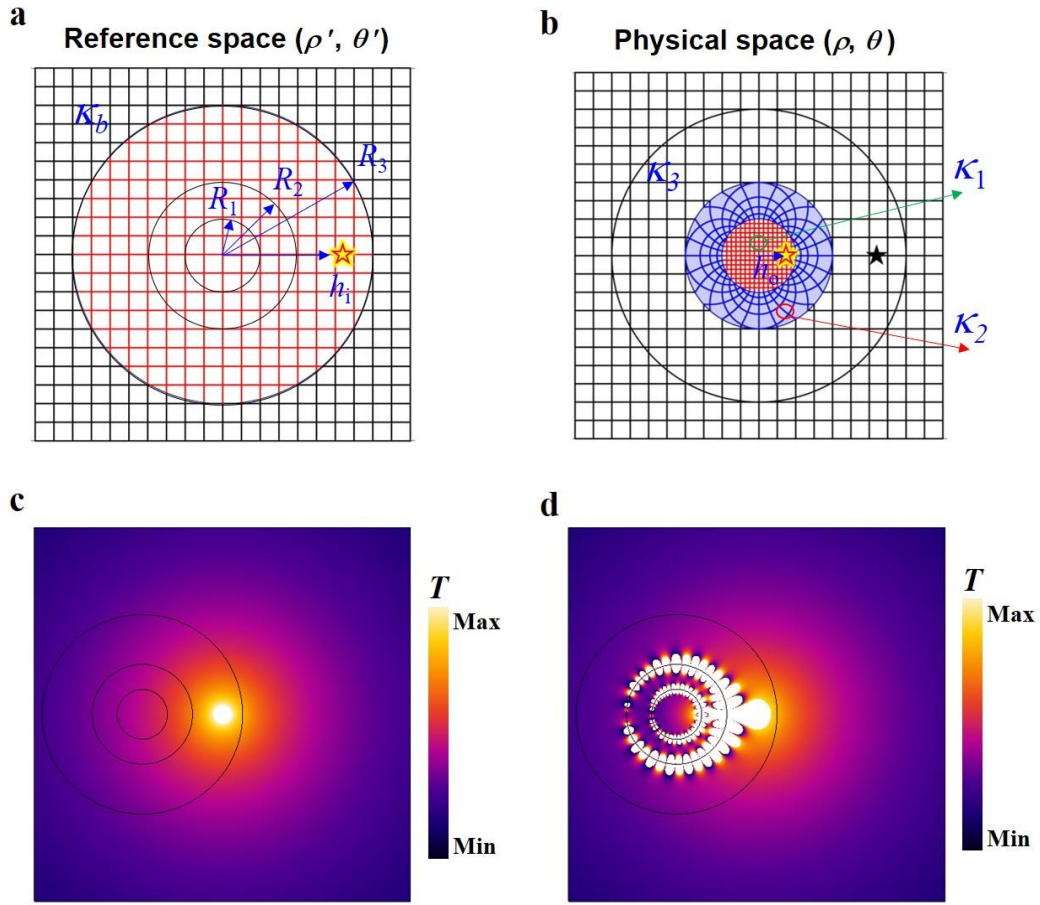

**Fig. 3| A folding transformation** between (**a**) the reference space and (**b**) the physical space, which is used to design a long-focus thermal lens with negative conductivity (blue region) that can create a virtual heat source (thermal image) at $f=h_i-R_2$ from the back surface of the thermal lens when a real heat source is at ($\rho=h_o$, $\theta = 0$) inside the lens. Simulated temperature distributions for (**c**) the reference space and (**d**) the physical space, which correspond to (**a**) and (**b**), respectively. The temperature distributions in (**c**) and (**d**) are almost the same (especially the region $\rho>R_3$), which means a line heat source at ($h_o$, 0) surrounded by the designed thermal lens with negative thermal conductive material can create almost the same temperature distribution as the case when a line heat source at ($h_i$, 0) with the same amplitude in a uniform positive thermal conductivity medium. The white regions mean that the temperatures are beyond the range of the color bar.

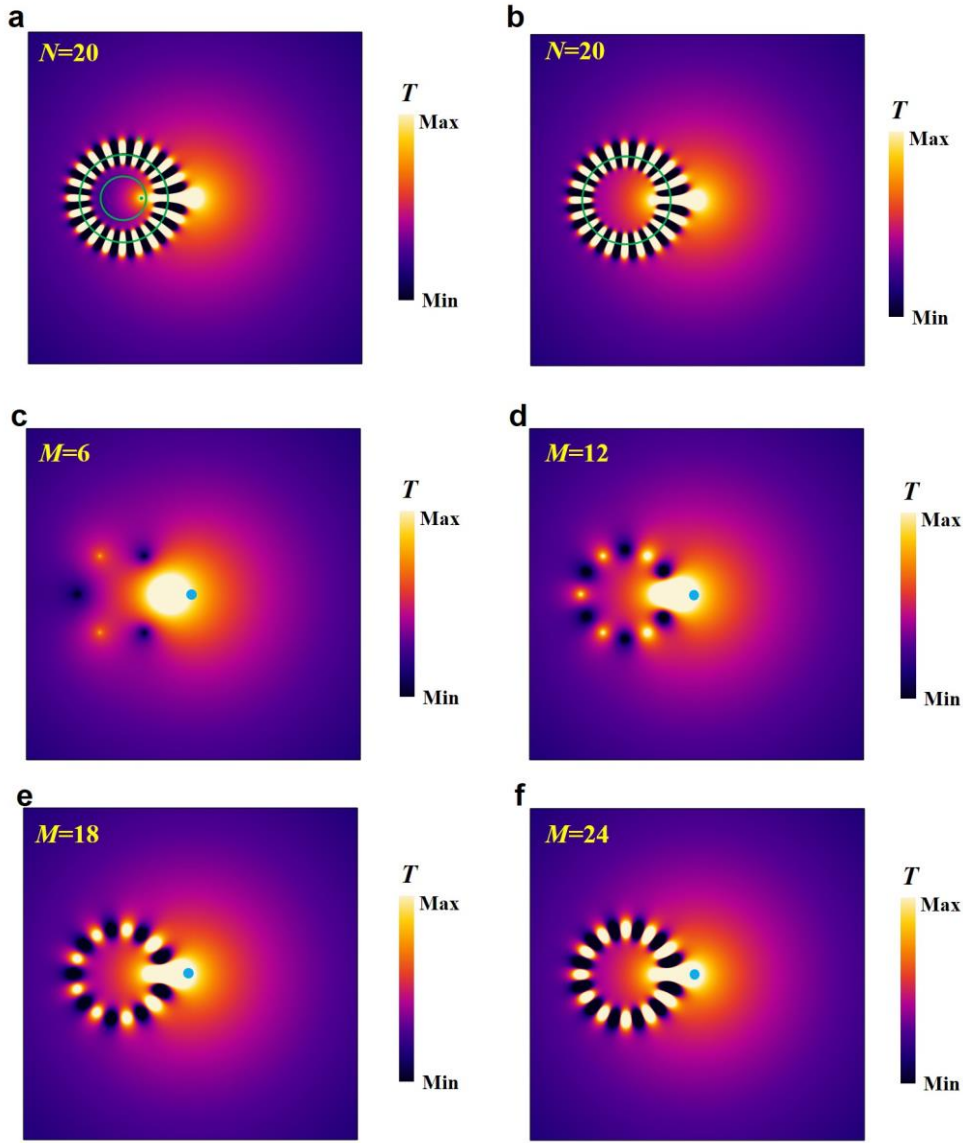

**Fig. 4| Simulated temperature fields** for (**a**) two-boundary heat sources along $C_1$ and $C_2$ given in Eq. (7) together with a line source at $\rho = h_o$, and (**b**) one-boundary heat source along $C_2$ in Eq. (8) without any line heat source. Green circles represent the location of the boundary sources. **c-f**, The simulated temperature fields for ATMS with different numbers along $C_2$, i.e., (**c**) $M = 6$, (**d**) $M = 12$, (**e**) $M = 18$, and (**f**) $M = 24$. The big blue dots indicate the position of the expected virtual source ($\rho = h_i$). Details of the numerical setting are given in Supplementary Note 5.

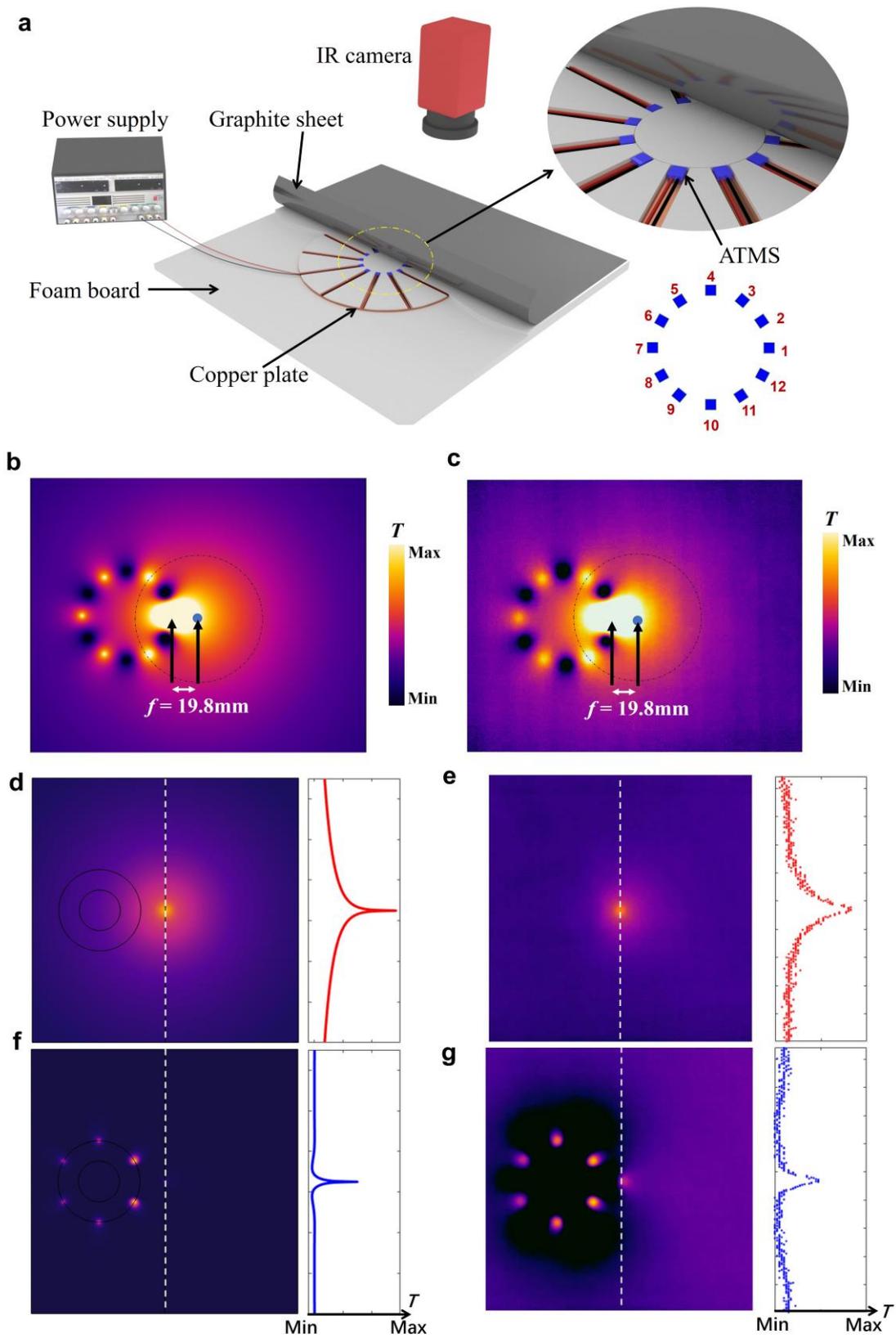

**Fig. 5| Experimental setup and measured results. a**, Schematic diagram of the designed thermal lens with 12 ATMSs to realize remote heating/cooling effect. The blue blocks are ATMSs whose serial numbers are marked in the illustration below it. The

brown circular region represents copper plate and the gray paper represents the graphite sheet. (**b**) and (**c**) are simulated and measured temperature distributions with the designed thermal lens to demonstrate the remote heating effect, respectively. As a reference to demonstrate the remote cooling effect, (**d**) and (**e**) are simulated and measured temperature distributions without the designed thermal lens, respectively, where only a line high-temperature source is in the background material. (**f**) and (**g**) are simulated and measured temperature distributions with the designed thermal lens for remote cooling effect, respectively, where a line high-temperature source is located at the focal point of the designed thermal lens.